\documentclass[epj]{svjour}
\usepackage{graphics}

\newcommand{\be}{\begin{equation}}
\newcommand{\ee}{\end{equation}}
\newcommand{\bea}{\vspace{0.25cm}\begin{eqnarray}}
\newcommand{\eea}{\end{eqnarray}}

\def\PLA{{Phys. Lett.}  A }
\def\PLB{{Phys. Lett.}  B }
\def\PRL{{Phys. Rev. Lett.} }
\def\PRA{{Phys. Rev.} A }

\def\PRD{{Phys. Rev.} D }

\begin{document}

%Title of paper
\title{ On the distances between entangled pseudoscalar mesons states. }

\author{M. Genovese }
\institute{Istituto Nazionale di Ricerca Metrologica, Strada delle
Cacce 91, 10135 Torino, Italy}

\date{}
\abstract{Entangled states of pseudoscalar mesons represent a very
interesting tool for studying foundations of quantum mechanics, e.g.
for testing Bell inequalities. Recently, they also emerged as a test
bench for quantum information protocols. On the other hand, from a
quantum information point of view, the characterization of the
distance between two quantum states is a topic of the utmost
importance. In this letter, with the purpose of providing a useful
tool for further investigations, we address the problem of which
distance allows a better discrimination between density matrices
appearing in pseudoscalar phenomenology.
 % end of PACS codes
} %end of abstract\begin{figure}
 \maketitle

\section{Introduction}

  Entangled states of neutral pseudoscalar mesons ($K_0$ and $B_0$)
represent a very interesting tool for studying foundations of
quantum mechanics \cite{prep}, e.g. as test of Bell inequalities
\cite{K,B,bram,did}. Recently, they also emerged as a test bench for
quantum information protocols \cite{QI}: for example protocols for
quantum teleportation \cite{tel} and quantum eraser \cite{eras} have
been proposed. These new proposals represent a very interesting
innovative application for $\Phi$ and B factories: indeed not only
to realize similar experiments with different physical systems is of
large interest, but also the use of mesons discloses new
possibilities \cite{prep,K,B,bram,did,QI,tel}.

On the other hand, from a quantum information point of view, a topic
of the utmost importance is the characterization of the distance
between two quantum states \cite{mat}, that, beyond its own
intrinsic interest, is needed, for example, for assessing
teleportation \cite{1}, purification \cite{2}, quantum cloning
\cite{3}, remote state preparation \cite{4} and state estimation
\cite{5}. The use of the neutral pseudoscalars in this context would
represent a further interesting application to quantum information.
Furthermore, the notion of distance between quantum states for kaons
is also  a useful tool when studying the developing of mixed states
from pure ones in presence of (quantum gravity induced) decoherence
\cite{qg} or when searching for CPT violation \cite{CPT}.

The problem of which distance is more convenient for distinguishing
different quantum states is not solved in general, but only specific
cases, as single qubits in a noisy channel \cite{mat}, have been
discussed.

In this letter we address this problem in  neutral pseudoscalar
mesons phenomenology. Several different experimental implementable
cases, which could represent an interesting "arena" for future
studies, are investigated. The presented results provide a useful
tool for future research addressed to study application of mesons
both to quantum information and decoherence, suggesting which
distance is better to use in specific cases.

\section{Distances between entangled pseudoscalar mesons states}

In the last years various distances among quantum states have been
defined \cite{karol} with the purpose of comparing the states and
eventually defining an entanglement measure.

Here we will consider three of the most used distances \cite{karol}:
the Bures, the Hilbert-Schmidt and the trace distances\footnote{In
the following we use the normalisations as in Ref. \cite{mat}.}.

The Bures distance, \be D_{B}(\rho , \sigma) =
\sqrt{1-(Tr[(\sqrt{(\sqrt{\sigma} \rho \sqrt{\sigma})}])^2} \ee
 represents the shortest path connecting two fibers (in the
Hilbert Schmidt fiber bundle) lying over the two density matrices.
It is both Riemannian and monotone \cite{karol}.

The Hilbert-Schmidt distance, \be D_{HS}(\rho , \sigma)
=\left(\frac{\sqrt{Tr[(\rho-\sigma).(\rho-\sigma)]}}{\sqrt{2}}\right)
\ee
 is the Euclidean distance deriving from the
definition of scalar product in the Hilbert-Schmidt space.  It is
Riemannian, but not monotone.

Finally the trace distance, \be D_{tr}(\rho , \sigma) = {1 \over 2}
Tr|\rho - \sigma| \ee simply derives by the definition of norm. It
is not Riemannian, but it is monotone. It coincides with the
Hilbert-Schmidt one for single qubits.

In the following we will consider the basis $K_L= \{1,0\}$,  $K_S=
\{0,1\}$, $K_L,K_S$ being the long and short living states of $K_0$
respectively (a similar discussion, \emph{mutatis mutandis}, for
mass and beauty eigenstates can be performed for B mesons as well
\cite{B}). The experimental determination of the meson states can be
achieved trough  decays \cite{K}.

The density matrix for the singlet state \footnote{in the following
the very small CP violation effects will be neglected, since they do
not substantially affect the presented results.} \be |\Psi^- \rangle
= { | K^0 \rangle | \bar K^0 \rangle - | \bar K^0 \rangle | K^0
\rangle \over \sqrt{2} } = { | K_L \rangle | K_S \rangle  - | K_S
\rangle | K_L \rangle \over \sqrt{2} } \ \ \ \, \label{psik} \ee
typically produced at $\Phi$ factories, is therefore:

\be \varrho_S = \left(
\begin{array}{llll}
 0 & 0 & 0 & 0 \\
 0 & \frac{1}{2} & -\frac{1}{2} & 0 \\
 0 & -\frac{1}{2} & \frac{1}{2} & 0 \\
 0 & 0 & 0 & 0
\end{array}
\right) \ee

As a first example, mathematically almost trivial, let us consider
the distances between the singlet and the transformed state after
that one of the two components has been regenerated in a slab
\cite{pred}, which is an interesting effect for a simple
experimental implementation.

The transformation is \cite{bram} ($f$ complex, with typical values
per unit thickness $|f| \sim 10^{-3}$ mm$^{-1}$ \cite{bram,did})

\be U = {1 \over \sqrt{(1 + |f|^2)}} \left(
\begin{array}{llll}
 1 & f & 0 & 0 \\
 f & 1 & 0 & 0 \\
 0 & 0 & 1 & f \\
 0 & 0 & f & 1
\end{array}
\right) \ee

 In this case the three distances coincide:
\bea D_{Bures}(\varrho_S, U \varrho_s U^{\dagger})=D_{tr}(\varrho_S,
U \varrho_S U^{\dagger})= \cr D_{HS}(\varrho_S, U \rho_s
U^{\dagger})=\sqrt{{|f|^2 \over 1 + |f|^2}} \eea

\begin{figure*}
\resizebox{0.75\textwidth}{!}{%
  \includegraphics{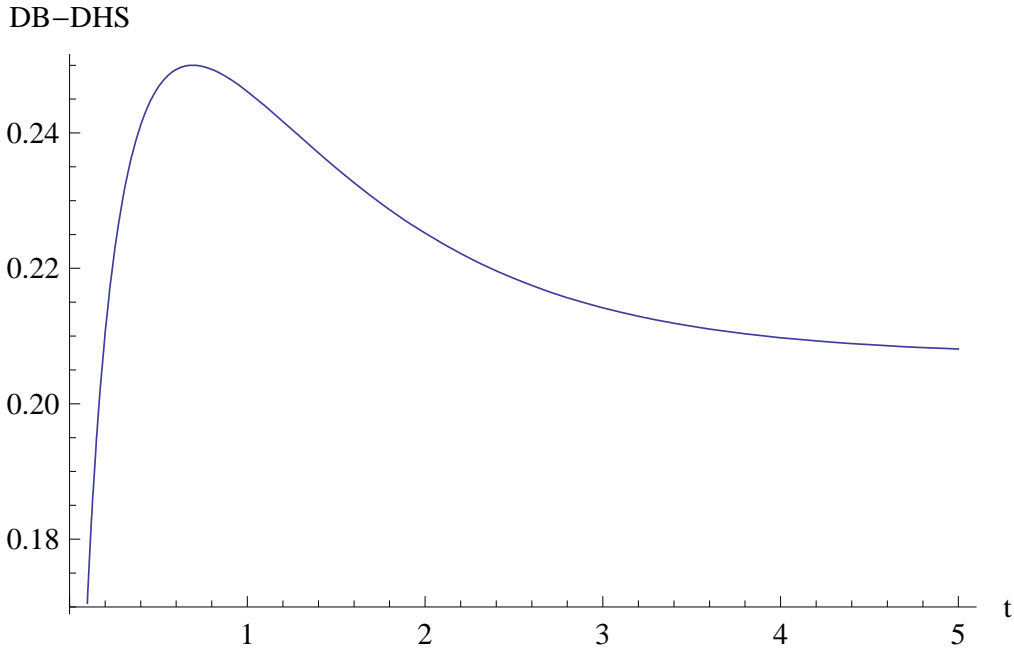}
}

\vspace*{5cm}       % Give the correct figure height in cm
\caption{Difference between Bures and Hilbert Schmidt distances
between singlet and decohered singlet in function of time (measured
in unities corresponding to decoherence characteristic time $1/l$) }
\label{1}       % Give a unique label
\end{figure*}

Let us now consider the more interesting case of the decoherence
effect on a singlet \cite{dec}, $t$ being the time \be
\left(\begin{array}{llll}
 0 & 0 & 0 & 0 \\
 0 & \frac{1}{2} & -\frac{1}{2} e^{-l t} & 0 \\
 0 & -\frac{1}{2} e^{-l t} & \frac{1}{2} & 0 \\
 0 & 0 & 0 & 0
\end{array}
\right) \ee This density matrix can describe both decoherence
effects due to environment interaction \cite{dec} and to gravity
induced decoherence \cite{qg}. Thus, one is interested in
distinguishing the decohered state from the unaffected singlet: a
point that has a large interest for future experiments.

Let us define $\tau=1/l$ the characteristic decoherence time scale.
Again the trace and Hilbert-Schmidt distances coincide. If one looks
to fig. \ref{1}, one can observe as the Bures distance allows a
better discrimination from the unaffected singlet at times smaller
than $\tau_s = 0.69 \tau$ , since it has a larger gradient. On the
other hand, the Hilbert - Schmidt one is more sensitive at larger
values of time. $\tau_s$ shifts toward smaller values if a
background is added, $\varsigma_D= x \sigma_D + (1-x) \textbf{1
}/4$, being $\tau_s = 0.67 \tau$ for a $1\%$ background and going to
$0.51$ for a $10 \%$ one.

As a further example, one can consider the statistical mixture of
the singlet with a singlet regenerated by a slab, $\varrho = x
\varrho_S + (1-x) U \varrho_S U^{\dagger}$, whose distance is
measured from the singlet. In this case the Hilbert-Schmidt
distance, coinciding with the trace one, is always less sensitive
than the Bures one.

The general case of the distance between the regenerated singlet
$U(f1) \varrho_S U(f1) ^{\dagger}$ and the mixing of the singlet
regenerated by two different slabs, $x U(f1) \varrho_S U(f1)
^{\dagger}+(1-x) U(f2) \varrho_S U(f2)^{\dagger}$ represents an
interesting example where a more complex situation can be studied
(and that eventually can also be implemented experimentally). This
case must be considered with a full variation of modulus and phase
of $f1,f2$. An example is reported in fig.2, where one can observe
the plot of the difference between Bures and Hilbert-Schmidt
distances when $f1,f2$ are real and $x=0.5$. For this case one can
see that, when varying $|f2|$ at fixed $|f1|$, the Hilbert Schmidt
distance is more sensitive than the Bures one when $|f2|<|f1|$ and
viceversa. When a phase is added to $f1$, the situation does not
change substantially, except for a decrease of the difference
$D_B-D_{HS}$ at small $f2$ (vanishing when the phase goes to $\pi /
2$) and a  small decrease of the inversion point under $|f1|$. On
the other hand, when we scan the variation with $x$ at fixed values
of $f1,f2$, the Bures distance is always more sensitive than the
Hilbert-Schmidt one when $x<0.75$ and vice-versa for $x>0.75$ (e.g.
see fig.\ref{3}).

\begin{figure*}
\resizebox{0.75\textwidth}{!}{%
  \includegraphics{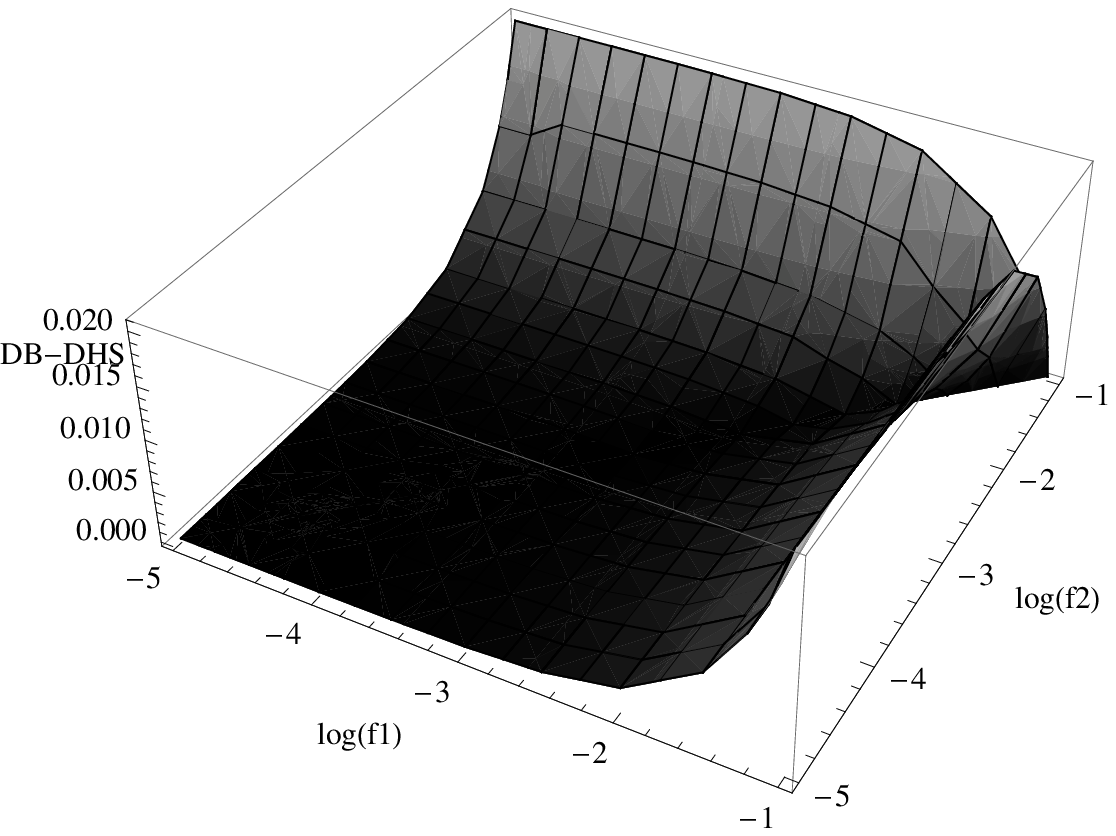}
}

\vspace*{5cm}       % Give the correct figure height in cm
\caption{3-dimensional plot of the difference between Bures and
Hilbert Schmidt distances
   between $U(f1) \varrho_S U(f1)^{\dagger}$ and $(U(f1) \varrho_S U(f1)^{\dagger} + U(f2) \varrho_S U(f2)^{\dagger})/2$
   in function of $log(f1)$ and $log(f2)$.}
\label{2}       % Give a unique label
\end{figure*}

\begin{figure*}
\resizebox{0.75\textwidth}{!}{%
  \includegraphics{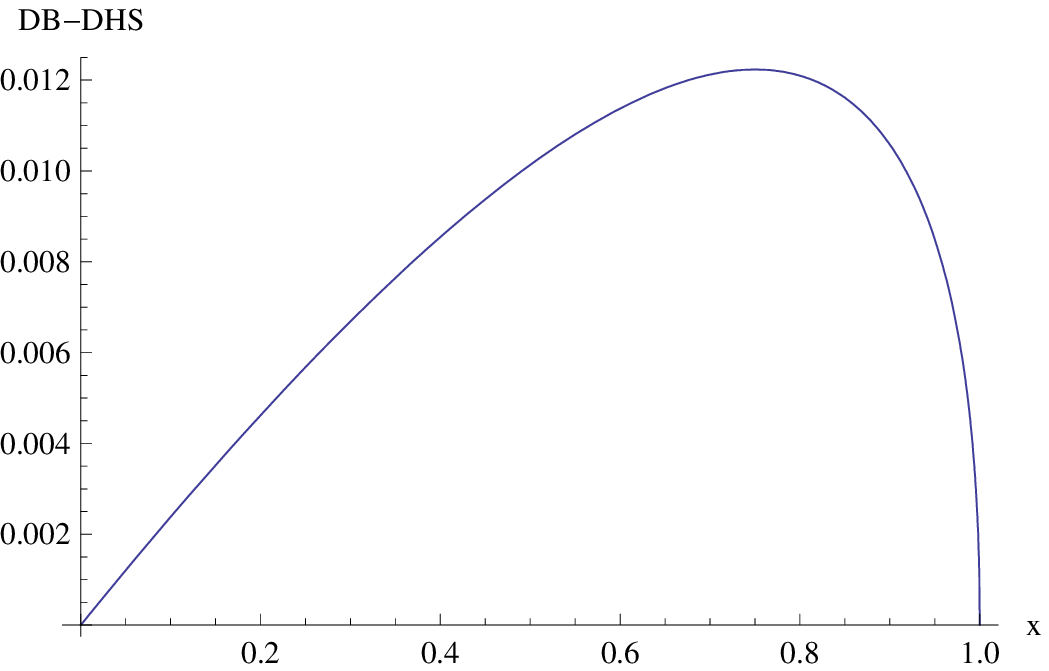}
}

\vspace*{5cm}       % Give the correct figure height in cm
\caption{Difference between Bures and Hilbert Schmidt distances
between
    $U(f1) \varrho_S U(f1)^{\dagger}$ and $x U(f1) \varrho_S U(f1)^{\dagger} + (1-x) U(f2) \varrho_S
    U(f2)^{\dagger}$,
  $f1=0.05,f2=0.001$, in function of $x$.}
\label{3}       % Give a unique label
\end{figure*}

In conclusion, we consider the case where the singlet is mixed with
a background, $\varrho = x \varrho_s + (1-x) \textbf{1 }/4$ [in
quantum information terminology a "depolarizing channel"]. Here, the
Hilbert - Schmidt distance is always more sensitive than the trace
one. On the other hand it is less sensitive than the Bures distance
for small  values of $x$ ($x < 0.5$), see fig.\ref{2}. Finally,
Bures distance is more discriminant than the trace one up to $x=2/3$
and less for larger $x$. The situation remains exactly the same when
considering the mixing $U \varrho_S U^{\dagger} + (1-x) \textbf{1 /
4}$ between regenerated singlet and background and changing the
phase and the modulus of $f$ (in a reasonable interval corresponding
to the experimental accessible zone \footnote{i.e. $|f|< 0.1$. For
$|f|$ approaching 1 the Hilbert-Schmidt distance becomes more
sensitive also at smaller $x$s.}).

\begin{figure*}
\resizebox{0.75\textwidth}{!}{%
  \includegraphics{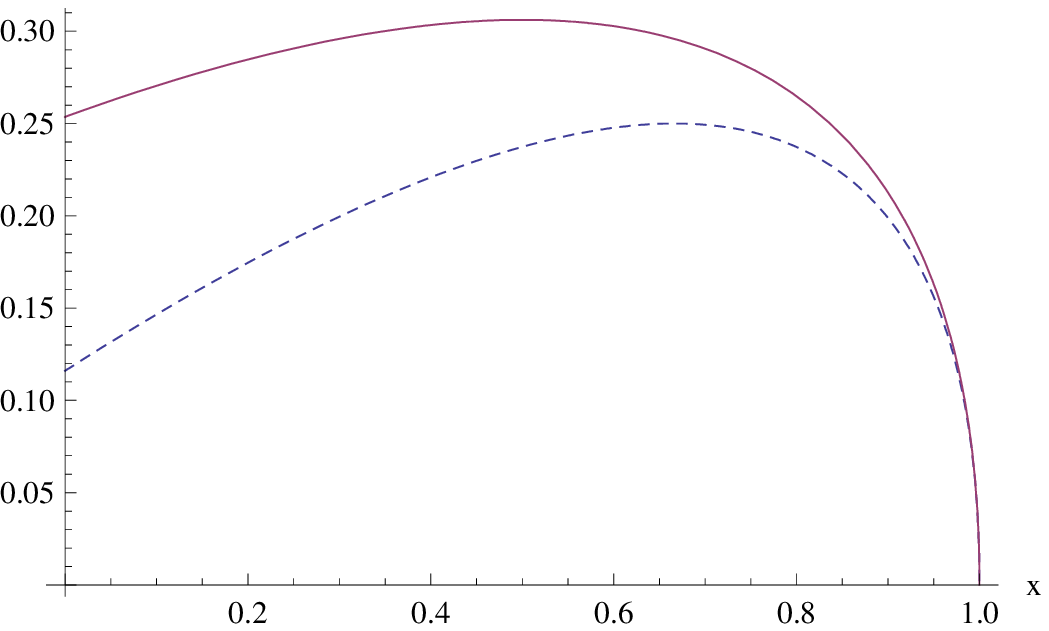}
}

\vspace*{5cm}       % Give the correct figure height in cm
\caption{Difference between Bures and Hilbert Schmidt distances
(solid) and between Bures and Trace
   distances (dashed)
   between singlet and singlet mixed with background, $\varrho = x \varrho_s + (1-x) \textbf{1}/4 $,
   in function of x.}
\label{4}       % Give a unique label
\end{figure*}

In summary, our calculations show that the choice of which distance
to consider for studying the $0^{-+}$ mesons entanglement depends on
the cases, in particular the Hilbert-Schmidt and the Bures ones look
to be the most promising. Our results suggest which would be the
best selection for some specific different situations of theoretical
and experimental interest.

\section{Conclusions}

In this letter we have presented a study addressed to estimate which
distance between states is more sensitive when comparing different
density matrices that can be met in neutral $0^{-+}$ mesons
phenomenology.

This is a relevant problem since neutral pseudoscalar mesons
represent an interesting test bench of quantum information
protocols, alternative to more traditional ones based on photons
and/or atoms. Many different interesting cases, from a quantum
information point of view,  can be experimentally achieved by
exploiting the regeneration phenomenon. Furthermore, our results can
find application when studying the developing of mixing from a pure
state due to gravitational induced (or not) decoherence.

Albeit our results are far from being exhaustive, nevertheless they
include many interesting examples pointing out how the sensitivity
of different distances varies from case to case and must be kept
carefully into account when applied to neutral pseudoscalar
phenomenology. In general, they suggest that Hilbert-Schmidt and
Bures distances are the most promising to be considered and they
represent a possible guide for a choice related to specific physical
examples, that could be  of use in further studies addressed to plan
meson experiments both on quantum information and decoherence.

\section{ Acknowledgements}

 This work has been supported by MIUR (PRIN
2005023443-002) and by Regione Piemonte (E14).

% Create the reference section using BibTeX:

\end{document}